\documentclass[
    ,final            
	,sort&compress,
  ]
  {aipproc}

\layoutstyle{8x11double}

\usepackage{amsmath,amssymb}
\usepackage{graphicx}

\newcommand{\e}{\mathrm{e}}

\newcommand{\bI}{\mathbf{I}}
\newcommand{\bS}{\mathbf{S}}

\newcommand{\bq}{\mathbf{q}}
\newcommand{\br}{\mathbf{r}}
\newcommand{\bx}{\mathbf{x}}
\newcommand{\by}{\mathbf{y}}
\newcommand{\mean}[1]{\langle #1 \rangle}

\begin{document}

\title{Magnetic Order in Kondo-Lattice Systems
due to Electron-Electron Interactions}


\author{Bernd Braunecker}{
  address={Department of Physics, University of Basel, 
             Klingelbergstrasse 82, 4056 Basel, Switzerland}
}

\author{Pascal Simon}{
  address={Department of Physics, University of Basel, 
             Klingelbergstrasse 82, 4056 Basel, Switzerland},
  altaddress={Laboratoire de Physique et Mod\'elisation des  Milieux
             Condens\'es, CNRS and Universit\'e Joseph Fourier, BP 166, 
             38042 Grenoble, France}
}

\author{Daniel Loss}{
  address={Department of Physics, University of Basel, 
             Klingelbergstrasse 82, 4056 Basel, Switzerland}
}

\begin{abstract}
The hyperfine interaction between the electron spin and the
nuclear spins is one of the main sources of decoherence for
spin qubits when the nuclear spins are disordered. An ordering
of the latter largely suppresses this source of decoherence.
Here we show that such an ordering can occur through 
a thermodynamic phase transition in two-dimensional (2D) 
Kondo-lattice type systems. We specifically focus on  
nuclear spins embedded in a 2D electron gas. 
The nuclear spins interact with each other through the RKKY interaction, which is carried 
by the electron gas. We show that a nuclear magnetic
order at finite temperature relies on the anomalous behavior 
of the 2D static electron spin susceptibility due to electron-electron interactions. 
This provides 
a connection between low-dimensional magnetism and non-analyticities
in interacting 2D electron systems.
We discuss the conditions for nuclear magnetism,
and show that the associated Curie temperature 
increases with the electron-electron interactions
and may reach up into the millikelvin regime.
The further reduction of dimensionality to one dimension is shortly discussed.
\end{abstract}

\maketitle

\section{Introduction}

A major source of decoherence of electron spin qubits is the hyperfine
coupling of the electron spin with the surrounding disordered nuclear spins \cite{burkard:1999a}. 
If we want to control and eventually eliminate this source of decoherence,
it is essential to fully understand the behavior of both the electron spin and the 
ensemble of nuclear spins. 
In this text we discuss such a fundamental aspect.
We address the question whether the nuclear spins can achieve order through 
a (ferro)magnetic phase transition. If they do the decoherence source
of the hyperfine interaction is massively reduced \cite{coish:2004a}.
We focus specifically on GaAs-based semiconductor heterostructures confining
two-dimensional electron gases (2DEGs). Such systems serve 
as the parent system for the single electron quantum dots defining
the spin qubits \cite{loss:1998}. 
Yet we emphasize that the described physics remains valid for 
general Kondo-lattice systems with interacting electrons.
It turns out that the anomalous properties of the 
2DEG, resulting from electron interactions, are crucial for the nuclear magnetic order.
The exposure to follow is an overview of our recent work published
in \cite{simon:2007,simon:2008}. 

Even though we shall be concerned mainly with fully translationally
invariant 2DEG, let us introduce the problem by considering
a single electron spin, confined to a quantum dot. 
The important interaction discussed here is the
hyperfine coupling between the electron spin on the
dot,
$\bS = (S^x, S^y, S^z)$, and the surrounding lattice of nuclear spins 
$\bI_i = (I_i^x, I_i^y, I_i^z)$ ($i$ is the index for the lattice site
at position $\br_i$).
The interaction Hamiltonian can be written in the form
\begin{equation} \label{eq:H_hyp_dot}
	H_{hyp}^{dot} = 
	\sum_i A_i \, \bS \cdot \bI_i
	=
	\sum_i A_i \, \bigl[ S^z I_i^z + S^+ I_i^- + S^- I_i^+ \bigr],
\end{equation}
with $S^\pm = S^x \pm i S^y$ and $I_i^\pm = I_i^x \pm I_i^y$,
and where $A_i \approx A |\psi(\br_i)|^2$ with $A$ a proportionality
constant and $\psi(\br_i)$ the wavefunction of the confined electron 
on the quantum dot.
The number of nuclear spins is large, typically of order
$10^{5}$, and so the nuclear spins generally act as a disordered bath 
on the electron spin.
While $I_i^z$ has a role similar to an external magnetic
field, the last two ``flip-flop'' terms in Eq. \eqref{eq:H_hyp_dot} flip the 
electron spin and lead to the decoherence.

This decoherence source can be largely suppressed when the nuclear spins
order, ferromagnetically or differently, which effectively suppresses the flip-flop 
terms in the Hamiltonian \cite{coish:2004a}.
There are two ways of polarizing the system, which we shall call the
\emph{extrinsic} (or dynamic) and the \emph{intrinsic} (or thermodynamic) polarization of the nuclear spin system.
The \emph{extrinsic} polarization consists in an active manipulation of the nuclear
spins by the experimentalist. Several methods to do this have been proposed 
and partially experimentally realized:
The development of 
quantum control techniques that effectively lessen or even suppress the nuclear
spin coupling to the electron spin \cite{johnson:2005,petta:2005,laird:2006};
the narrowing of the nuclear spin distribution \cite{coish:2004a,Klauser,Stepanenko};
or the dynamical polarization of the nuclear 
spins \cite{burkard:1999a,khaetskii:2003,imamoglu:2003,bracker:2005a,coish:2004a}.
Yet in order to 
extend the spin decay time  by one order of magnitude through polarization 
of the nuclear spins, a polarization of  above 99\% is required \cite{coish:2004a},
quite far from 
the best result reached to date in quantum dots, which is about 60\% \cite{bracker:2005a}.

It is possible, however, that a full polarization is achieved \emph{intrinsically} as well,
i.e. through a thermodynamic phase transition to, for instance, a ferromagnetic state.
This is our main topic here.
In what follows we give a qualitative, physical account to this possibility
by introducing step by step the model, the necessary conditions, and the results. 
For details we refer to \cite{simon:2007,simon:2008}.
We first indicate how to obtain from a microscopic model an effective 
Hamiltonian for the nuclear spins only. 
We show that the conditions of the Mermin-Wagner theorem are not 
met so that long range order is not forbidden in the 2D system.
With a simple mean field theory we can then see under which conditions ferromagnetic
order is possible and allows us to identify the first important 
temperature scale $T_{MF}$. A refinement leads to the second important
temperature scale $T^*$, which depends on the electron-electron interactions
in the 2DEG. The calculation of the susceptibility for interacting electrons
allows to estimate $T^*$ and on the stability of the 
magnetic order. It turns out that nonanalytic corrections to the 
electron spin susceptibility are crucial. We discuss the possible forms of
these corrections, allowing us to conclude with some numerical
estimates for a possible nuclear order.


\section{Model and effective model}
 
In order to discuss thermodynamics we have to shift our point of view
from electrons in quantum dots to a fully translationally invariant
2DEG as it can be obtained,
for instance, in a GaAs heterostructure. 
The Hamiltonian of such a system can be written as 
\begin{equation} \label{eq:H}
	H = H_{el} + H_{hyp},
\end{equation}
where $H_{el}$ describes the Hamiltonian of the interacting 
electron gas, and 
\begin{equation}
	H_{hyp} = \sum_i A \bS_i \cdot \bI_i
\end{equation}
is the hyperfine interaction between the electron and nuclear spins.
We have chosen here a tight binding, Kondo-lattice formulation of the problem, 
where $\bS_i = (S_i^x,S_i^y,S_i^z)$
is the operator of an electron spin in a Wannier state centered at lattice
site $i$ and the $\bI_i = (I_i^x,I_i^y,I_i^z)$ 
are the nuclear spin operators as before.
For GaAs we have $I= 3/2$.
Notice that $A$ now is position-independent due to the translational
symmetry. 

In Eq. \eqref{eq:H} we have not included the direct dipolar interaction
between the nuclear spins. This interaction has the smallest energy scale in
the system, $E_{dd} \approx 100$ nK \cite{paget:1977}. 
It is much weaker than the effective nuclear
spin interaction discussed below and, in particular, $E_{dd}$ is much smaller
than typical experimental temperatures. This allows us to entirely neglect 
the direct dipolar interaction.

For GaAs, $A \approx 90$ $\mu$eV \cite{paget:1977}, which compares to typical
Fermi energies of $E_F \approx 10$ meV \cite{beenakker:1991}. 
The small ratio $A/E_F \sim 10^{-2}$ implies a separation of time scales. 
Electron relaxation times are much shorter than typical time scales of 
the nuclear spins. This allows us to decouple the systems and focus
on the magnetic properties of the nuclear spin system alone, where 
the effective spin-spin interactions are carried through the 
response of the equilibrium electron gas to local magnetic
excitations, i.e. the electron spin susceptibility.

Technically, the first step is to reduce the still quasi-2D (due to the finite
thickness of the 2DEG) problem to a true 2D problem. Since the electrons are 
confined in a single mode in the direction orthogonal to the 2D plane, 
the nuclear spins along a column in this direction are all almost
identically coupled to the electrons and so effectively locked in a 
ferromagnetic alignment and behave like a single (effectively large)
nuclear spin. The problem becomes, therefore, truly 2D.

The effective Hamiltonian describing this situation can then be obtained,
for instance, through a Schrieffer-Wolff transformation 
followed by the integration over the electron degrees of freedom
\cite{simon:2007,simon:2008}. It is given by 
\begin{equation} \label{eq:Heff}
	H_\text{eff} = 
	- \sum_{ij \alpha} J^{\alpha}(\br_i - \br_j) I_i^\alpha I_j^\alpha
	=
	- \frac{1}{N} \sum_{\bq \alpha} J_{\bq}^{\alpha} I_{-\bq}^\alpha I_\bq^\alpha,
\end{equation}
where $N$ is the number of sites in the system, $\alpha = x,y,z$ and 
the lattice indices $i,j$ run over the two-dimensional lattice
with site positions $\br_i, \br_j$.
Furthermore
\begin{equation}
	J^{\alpha}(\br_i-\br_j) =
	- \frac{A^2}{8n_s} \chi^{\alpha}(\br_i-\br_j)
\end{equation}
is the effective Ruderman-Kittel-Kasuya-Yosida (RKKY)
interaction \cite{RKKY}, with $n_s = a^{-2}$ the nuclear spin density
and 
\begin{equation}
	\chi^{\alpha}(\br_i-\br_j)
	= -\frac{i}{\hbar}
	\int_0^\infty \mathrm{d}t \ \mean{[S_i^\alpha(t) \, , \, S_j^\alpha(0)]}
	\e^{-\eta t},
\end{equation}
is the static electron spin susceptibility ($\eta>0$ is infinitesimal). 
The Fourier transforms are defined as 
$I_\bq^\alpha = \sum_i \e^{i \br_i \cdot \bq} I_i^\alpha$ and
$J_\bq^\alpha = \int \mathrm{d}\br \, \e^{-i \br \, \cdot \bq} J^\alpha(\br)$,
and $N$ is the number of sites in the system.
We shall henceforth consider only isotropic electron systems, allowing us 
to drop the $\alpha$ index in $\chi$ and $J$.

Much of the magnetic properties of the nuclear spins depends, therefore,
directly on the shape of $J_\bq$, i.e. on the electron susceptibility $\chi(\bq)$.
Below we will first investigate which features of $\chi(\bq)$ are required
such that nuclear ferromagnetism is stable. Then we shall see that 
electron interactions can indeed lead to such a behavior.

Yet before starting we have to comment on the Mermin-Wagner theorem \cite{MW}, which 
states that long range order in Heisenberg-like models in 2D is
impossible, provided that the interactions are sufficiently short ranged.
The RKKY interaction, however, is long ranged. But it is also oscillatory,
and it has been conjectured recently \cite{bruno:2001} that the Mermin-Wagner
theorem extends to RKKY interactions that are carried by noninteracting 
electrons. Below we find a direct confirmation of this conjecture, indicating that
electron interactions in addition to the long range character of $J_\bq$
play the crucial role.


\section{Mean field theory}

As a first (naive) approach, we can look at the problem of nuclear
magnetism on the mean field level, similar to the approach used by 
Fr\"ohlich and Nabarro (FN) for bulk metals 
more than 60 years ago \cite{froehlich:1940}. 
We skip here the explicit calculation as it is a standard Weiss mean 
field calculation. Instead, let us clearly state the main assumption
behind FN's approach: If we look at the Hamiltonian
\eqref{eq:Heff} we see that the energy can classically be minimized
if the $I^\alpha_i$ align in a single spatial Fourier mode $\bq$ corresponding
to the maximum of $J_\bq$. If this maximum is reached at $\bq = 0$,
the ground state is ferromagnetic. 
FN implicitly assumed that the physics is entirely determined by this
maximum energy scale $J_0$ and investigated the Hamiltonian
$H_{FN} = - \frac{J_0}{N} \sum_{ij} \bI_i \cdot \bI_j$, for which 
the mean field theory is exact. 

We have to retain two important points from this theory: 
On the one hand, the ground state depends on the maximum of the $J_\bq$,
and a ferromagnet is only possible if it is reached at $\bq=0$.
A $\bq \neq 0$ implies a different magnetic order, e.g. a helimagnet.
On the other hand, this theory depends on a single energy scale,
$\max_{\bq} J_\bq$ (e.g. $J_0$), which we associate with a temperature $T_{MF}$ 
through 
\begin{equation}
	k_B T_{MF} = \max_{\bq} J_\bq,
\end{equation}
where $k_B$ is the Boltzmann constant. Since this is the only energy
scale in the system, all thermodynamic quantities must directly depend
on it. For instance, the Curie temperature is given by \cite{froehlich:1940}
\begin{equation} \label{eq:Tc_MF}
	T_{c}^{MF} = \frac{I(I+1)}{3} T_{MF}.
\end{equation}


\section{Refinement}

\begin{figure}[t]
	\includegraphics[width=\columnwidth]{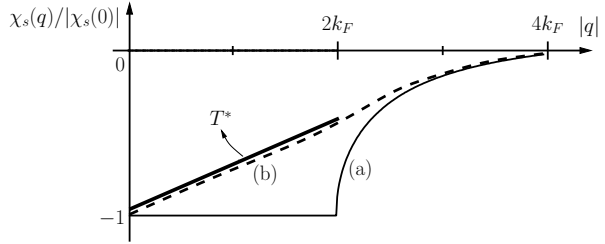}
	\caption{
	Electron spin susceptibility $\chi(q)$ for 
	(a) noninteracting electrons in 2D (Lindhard function),
	(b) interacting electrons such that a nuclear ferromagnet
	would be stable. In case (b) the interactions introduce a new
	scale $T^*$ associated with the slope of the curve at small $q$
	[see Eq. \eqref{eq:T*}], represented by the thick line next to
	the curve.
	\label{fig:chi}
	}
\end{figure}

The mean field theory is, however, inconsistent with the RKKY interactions
for noninteracting electrons. In the noninteracting case the electron spin susceptibility 
is exactly known and is given by the Lindhard function \cite{GV},
which is constant for $0 < |\bq| < 2k_F$
[see Fig. \ref{fig:chi} (a)], where $k_F$ is the Fermi momentum.
There is therefore no well defined $\bq$
	at which $J_\bq$ is maximum. ``Guessing'' a ground state about which we
can proceed with a mean field theory as above is no longer possible. 

\begin{figure}[t]
	\includegraphics[width=\columnwidth]{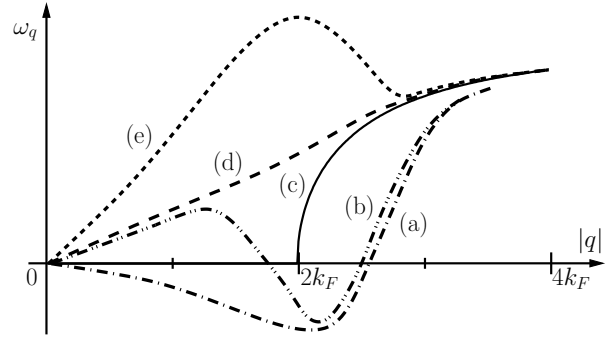}
	\caption{
	Possible shapes for the magnon dispersion $\omega_q$ [Eq. \eqref{eq:omega_q}],
	equal to the shifted and rescaled electron spin susceptibility 
	$\chi(q)$.
	Noninteracting electrons lead to the curve (c), obtained from
	the Lindhard function. 
	Electron interactions modify the shape of this curve and can in principle lead
	to any of the curves (a)--(e).
	Ferromagnetic order is unstable in cases (a) and (b)
	where the magnon dispersion would become negative; 
	a helical magnetic order becomes here possible.
	Nuclear magnetism is unstable 
	in case (c), which has a continuum of excitations with zero energy. 
	A nuclear ferromagnet is stable for cases (d) and (e).
	\label{fig:omega}
	}
\end{figure}

In fact, it is straightforward to show that such a ground state 
cannot be stable. For this purpose we look at fluctuations about 
an ordered ground state. In Heisenberg type systems the lowest lying excitations
are magnons, collective long ranged spin wave excitations.
The calculation of the magnon dispersion about a ferromagnetic order
is a standard textbook exercise (see e.g. \cite{ashcroft}) and leads to
\begin{equation} \label{eq:omega_q}
	\hbar \omega_\bq = 2 I (J_0 - J_\bq).
\end{equation}
We shall henceforth assume that $J_\bq$ is independent of the direction 
of $\bq$ and write $J_q$, $\chi(q)$, as well as $\omega_q$.
The fact that $J_q$ is constant for $0 < q < 2k_F$ means that there
is a continuum of magnon excitations at zero energy $\omega_q = 0$
[see Fig. \ref{fig:omega} (c)],
and every magnon slightly decreases the global magnetization.
The assumed ground state magnetization cannot be stable, the nuclear 
spin system
is disordered. This is a direct illustration of the extension of 
the Mermin-Wagner theorem conjectured in \cite{bruno:2001}.

From Eq. \eqref{eq:omega_q} we see, however, that a ferromagnetic
ground state becomes stable if $\omega_q$ 
[i.e. $\chi(q)$] increases monotonically
with $q$ [Fig. \ref{fig:chi} (b) and Fig. \ref{fig:omega} (d),(e)]. 
As we shall see below, nonanalytic 
corrections to $\chi(q)$ by electron-electron interactions can indeed
lead to a \emph{linear} increase in $q$. 
The main effect is illustrated in Fig. \ref{fig:chi} (b). 
The electron-electron interactions modify the RKKY interaction $J_q$
in that they introduce a new energy scale $T^*$, which characterizes
the shape of $J_q$.
Stronger electron-electron interactions lead to a larger slope of the 
linear increase of $\chi(q)$ at small $q$. Using this slope
and $2k_F$ as the only inverse length scale available for the 
electron gas, the new energy scale must be set by the quantity
\begin{equation} \label{eq:T*}
	k_B T^* = (2k_F) \left.\frac{\mathrm{d} J_q}{\mathrm{d}q}\right|_{q=0}.
\end{equation}
The physics is, therefore, no longer dominated by the single scale $T_{MF}$. 
Thermodynamic quantities such as the magnetization per site $m(T)$ or
the critical temperature $T_c$ must be 
a function of these available scales: 
$m(T)=m(T_{MF},T^*;T), T_c = T_c(T_{MF},T^*)$, etc.

We can again use the magnon description to shed more light on this
dependence. 
The average magnetization per site can be written as
\begin{equation}
	m 
	= I - \frac{1}{N} \sum_\bq \frac{1}{\e^{\hbar \omega_q / k_B T}-1}
	= I - \frac{a}{2\pi} \int \frac{\mathrm{d}q \, q}{\e^{\hbar \omega_q / k_B T}-1},
\end{equation}
with $a$ the nuclear lattice constant. The summation/integration runs
over the first Brillouin zone of the nuclear system. We see that
this integral converges if $\omega_q \propto q$ for $q \to 0$, and 
so the linear corrections to the susceptibility are essential for the 
existence of a finite critical temperature. 
The magnon integral is dominated by this linear behavior up to $T \sim T^*$.
For these temperatures, we can explicitly calculate the magnon integral
and obtain
\begin{equation} \label{eq:m}
	m = I \left(1- T^2/T_0^2\right),
\end{equation}
with
\begin{equation}
	T_0 = \frac{I}{2k_F} \sqrt{\frac{3I n_s}{\pi}} T^* \sim \frac{\lambda_F}{a} T^*,
\end{equation}
where $\lambda_F= 2\pi/k_F$ is the Fermi wavelength.  

We stress that the calculation leading to Eq. \eqref{eq:m}
is valid only in the limit $T\lesssim T^*$, where the spin waves form 
a dilute gas. At higher temperatures the number of spin waves increases
and their wavelengths become shorter, leading to a breakdown of 
the spin wave theory. The temperature $T_0$ in Eq. \eqref{eq:m} has
to be interpreted accordingly: We see that $T_0$ sets a characteristic
temperature scale for the magnetization $m(T)$. 
Further corrections to Eq. \eqref{eq:m}, and mainly the dependence on
$T_{MF}$, are of exponential form, $\sim \e^{- T_{MF}/T}$, and are
uniquely determined by the spin wave modes beyond the validity of the theory. 
In addition, from Fig. \ref{fig:chi}, we see that $T^*$ generally
(i.e. for not too weak interactions)
is comparable to $T_{MF}$, which is proportional to the maximum of $\chi(q)$.
This means that up to temperatures $T \sim T_{MF}$, the magnetization 
is essentially independent of the $T_{MF}$ scale.

If we conjecture on this basis that $T_{MF}$ indeed is insignificant
for the thermodynamics, then $T_0$ must set the scale for the 
Curie Temperature $T_c$,
\begin{equation}
	T_c \sim T_0 \sim \frac{\lambda_F}{a} T^*.
\end{equation}
In contrast to the mean field result \eqref{eq:Tc_MF}, 
this estimate is consistent with $T_c \to 0$ for noninteracting electrons. 
The prefactor $\lambda_F / a \sim 10^{2}$ (in GaAs) is a consequence
of coupling the electron system (with length scale $\lambda_F$) to the
nuclear spin system (with length scale $a$), and leads to a strong
increase of the characteristic temperature $T_0$ compared with $T^*$. 
A further discussion of the renormalization of $T_0$ can be found
in \cite{simon:2008}.

In order to give a numerical estimate of $T_0$ we need to investigate
the interactions between electrons in the 2DEG.
%


\section{Interaction corrections to the susceptibility}

Interactions and correlations between particles become increasingly
important in low dimensions due to a reduction of the available
phase space for particle scattering. 
It is not too surprising, therefore, that electron-electron interactions
in the 2DEG lead to deviations from the thermodynamics of standard
Fermi liquids. Such deviations have attracted some attention over the 
last years theoretically 
\cite{belitz:1997,hirashima:1998,misawa:1998,chitov:2001,maslov:2003,
maslov:2006,millis:2006,efetov:2006,shekhter1:2006,shekhter2:2006,chubukov:2007} and 
experimentally \cite{reznikov:2003}. 
Of specific importance for our case are Refs. \cite{maslov:2003,maslov:2006,millis:2006},
where explicitly self-energy corrections to the susceptibility were calculated for a screened
Coulomb interaction $U(\bx-\by) = U \delta(\bx-\by)$ in 2D. 
Nonanalytic behavior appears at second order perturbation in $U$ and leads to 
\begin{equation} \label{eq:chi_CM}
	\delta\chi(q) 
	= 
	\chi(q)-\chi(0)
	= -  q \frac{4 |\chi(0)| |\Gamma_s|^2}{3\pi k_F},
	\quad q \ll 2k_F,
\end{equation}
where $\Gamma_s = - U m / 4\pi$ is the bare $2k_F$ backscattering 
vertex and $m$ the effective mass.
We emphasize the nonanalytic
behavior on the modulus $q=|\bq|$, which cannot be derived within a 
standard Fermi liquid theory. 
This linear $|\bq|$ dependence is indeed the necessary dependence for 
a nuclear ferromagnet as discussed above. 
The problem here is, however, the sign of this correction: 
If we feed this $\chi(q)$ back into the spin wave spectrum \eqref{eq:omega_q},
we see that $\omega_q$ becomes \emph{negative} for $q \neq 0$, meaning that
the assumption of ferromagnetic order is incorrect and that such an order
is in fact unstable.
In such a situation a different, helical order can be possible,
and this interesting possibility is discussed in \cite{simon:2008}. 
Here we focus instead on a further renormalization
of the nonanalytic correction, which can reestablish the ferromagnetic order.

We restate the comment made at the beginning of this section that 
correlation effects are important in low dimensions. A result from perturbation 
theory, such as Eq. \eqref{eq:chi_CM}, undergoes further renormalization 
by higher order processes and can so change considerably its shape.
This indeed can happen if we push further the diagrammatic 
calculation of \cite{maslov:2003,maslov:2006,millis:2006} and 
include the full summation of selected classes of diagrams such as,
for instance, the Cooper channel renormalization of the two-particle scattering
vertex \cite{saraga:2005}.
The Cooper channel renormalization for $\chi(T)$ has been considered
recently in \cite{efetov:2006,shekhter1:2006,shekhter2:2006,chubukov:2007}.
The effect of this renormalization on $\chi(q)$ 
has been estimated in \cite{simon:2008} and very recently explicitly
calculated in \cite{chesi:2008}. The obtained corrections to $\chi(q)$ are
nonuniversal and depend on detailed cutoff scales of the 
renormalization such as Fermi energy, temperature, level spacing, etc. 
The stability of the ferromagnetic (or helical) phase seems, therefore, to depend on 
a quantity which is difficult to control. 

Yet we must note that
this is the result of a perturbative renormalization, the summation over 
selected classes of diagrams, as well as the result of a screened short ranged
electron-electron interaction. Alternative approaches lead to more 
predictable results: If we consider long ranged Coulomb interactions
within a local field factor approximation (which is a semi-empiric generalization 
of the RPA approximation \cite{GV}), the spin wave spectrum is 
always positive and a ferromagnet is stable. 
Such effective theories, however, have eventually the same difficulty 
of control as the summation over classes of diagrams. 
Due to this, it seems so far that the final determination
of the shape of the susceptibility probably has to rely on experiments
and numerics. The latter two approaches should then probe not only 
the slope $\partial \chi/\partial q$ at $q \approx 0$ but also 
$\chi(q \sim 2k_F)$, which is important to discriminate
between the different scenarios shown in Fig. \ref{fig:omega}
by the curves (a) and (b) on the one
hand, and by (e) and (f) on the other hand.
We stress that numerics should directly target
$\chi(q)$ and not, as for instance done in \cite{moroni:1995,davoudi:2001}, 
the local field factor. The relation between the latter and 
$\chi(q)$ is actually singular at $q=2k_F$, which amplifies the 
noise in the Monte Carlo data of \cite{moroni:1995,davoudi:2001} 
and makes a conclusion on $\chi(q)$ unreliable.

Assuming, however, that we have found a window in which 
a nuclear ferromagnet is stable, we can
estimate $T_0$ from the different calculation schemes.
A detailed discussion is given in \cite{simon:2008}. 
We find that remarkably all schemes provide comparable values. 
As anticipated much depends on the strength of the electron
interactions, which can be quantified by the commonly used
dimensionless parameter $r_s$ (see e.g. \cite{GV}) 
expressing roughly the ratio 
of Coulomb over kinetic energies of the electrons.
In the 2DEG $r_s$ scales with the electron density $n_e$ and $k_F$
as $r_s \propto 1/\sqrt{n_e} \propto k_F^{-1}$. 
Values up to $r_s \sim 8$ can be reached experimentally nowadays.
With increasing $r_s$, the scale $T_0$ is enhanced through two main 
effects:
First, $k_F^{-1}$ increases linearly with $r_s$.
Second, $|\chi(0)|$, which is essentially the Pauli susceptibility
at small $r_s$, increases linearly with $r_s$ \cite{GV}.
We moreover note that larger $r_s$ drive the system closer
to the ferromagnetic Stoner instability,
which would occur at $r_s \sim 20$ \cite{senatore:2001}.
At this instability $\chi(0)$ would diverge. The proximity 
to the instability provides an additional prefactor enhancing
$|\chi(0)|$.
Let us note that the increase of $|\chi(0)|$
enhances both, $T^*$ and $T_{MF}$.
For $r_s \sim 5$, we then obtain $T_0 \sim 0.3 - 0.4$ mK, and
for $r_s \sim 8$ the larger $T_0 \sim 0.7 - 1$ mK, 
where the spread is due to the different analytic approaches
\cite{simon:2008}.


\section{One-dimensional systems}

With a further reduction of dimensionality electron-electron
correlations become even more important. 
One-dimensional conductors of interacting electrons, such as quantum wires or carbon 
nanotubes, form a Luttinger liquid rather than a Fermi liquid.
Accordingly the shape of the RKKY interaction $J_q$
between the nuclear spins in these systems changes drastically.
$J_q$ is dominated by backscattering processes at $q = 2k_F$, and the
nuclear spins order in a helical phase with this wave vector. 

In contrast to the 2D case, the feedback of the ordered nuclear 
field on the electrons is now crucial. It leads to a spontaneous restructuring
of the electron wave functions, \emph{i.e.}\/, to an order in the electron system
as well.
The feedback stabilizes this order, and the 
critical temperature $T_c$ can increase by 
several orders of magnitude compared with the case where the 
feedback has been neglected.

We have performed a detailed study of this effect 
in \cite{braunecker:2008} for the 
example of single-wall carbon nanotubes made from the $^{13}$C isotope (which
has a nuclear spin $I=1/2$). Such nanotubes have become
available very recently \cite{marcus:2008,simonf:2005,ruemmeli:2007}.
The hyperfine interaction is very weak in such systems. 
Due to the feedback, however,
we determine a $T_c$ in the millikelvin range. The ordered
phase leads furthermore to a universal reduction of the conductance
and should be detectable by standard transport measurements.


\section{Conclusions}

We have discussed here the conditions necessary for ferromagnetism
of nuclear spins embedded in a 2DEG. Such a system is naturally in the 
RKKY regime and the interaction between the nuclear spins is carried by
the electron spin susceptibility. 
Electron correlations are crucial, and the stability of the magnetic
(ferro or different) phase depends on the nonanalytic behavior, linear in momentum $|\bq|$,
of the electron spin susceptibility. 
The nonanalytic behavior cannot be found from standard Fermi liquid theory. 
It is a consequence of a strong renormalization of the electron-electron 
interaction which, to the best of our knowledge, strongly depends on 
a nonuniversal cutoff  scale specific for the sample under investigation.
The stronger the interaction, however, the higher also the critical 
temperature for the nuclear (ferro)magnet.
Our estimates show that a transition temperature reaching up into the millikelvin 
range may be achievable. 
A similar temperature range is estimated for one-dimensional 
conductors such as the recently available $^{13}$C single-wall nanotubes,
where the ordered phase should be detectable by conductance measurements.

To conclude, we stress that 
such physics is not restricted to nuclear spins in metals. We 
expect a similar behavior for any Kondo-lattice system in low dimensions
with interacting conduction electrons.

\begin{theacknowledgments}
We are grateful to D. Maslov for useful discussions.
This work was supported by the Swiss NSF, NCCR Nanoscience, 
and JST ICORP.
\end{theacknowledgments}



\end{document}